\documentclass[useAMS,usenatbib,usegraphicx]{mn2e}

\usepackage{graphicx}
\usepackage[latin1]{inputenc}
\usepackage{color}
\usepackage{times}
\usepackage{natbib}
\usepackage{setspace}
\newif\ifAMStwofonts
\AMStwofontstrue
\definecolor{red}{rgb}{1,0.,0.}

\newcommand{\morgana}{{\sc morgana }}
\newcommand{\munich}{{\it Munich }}
\newcommand{\somer}{SC-SAM }

\def\lesssim{\lower.5ex\hbox{$\; \buildrel < \over \sim \;$}}
\def\gtrsim{\lower.5ex\hbox{$\; \buildrel > \over \sim \;$}}
\title[Mergers and Morphologies] {On the dependence of galaxy
  morphologies on galaxy mergers}

\author[Fontanot et al.]{
  \parbox[t]{\textwidth}{Fabio Fontanot$^{1,2}$\thanks{E-mail:
      fontanot@oats.inaf.it}, Andrea V. Macci\`o$^3$, Michaela
    Hirschmann$^4$, Gabriella De Lucia$^1$, Rahul Kannan$^{5,3}$,
    Rachel S. Somerville$^6$, Dave Wilman$^{7,8}$}
    \vspace*{8pt}\\
    $^1$ INAF - Astronomical Observatory of Trieste, via G.B. Tiepolo 11, I-34143 Trieste, Italy \\
    $^2$ HITS - Heidelberger Institut f\"ur Theoretische Studien, Schloss-Wolfsbrunnenweg 35, 69118 Heidelberg, Germany\\
    $^3$ Max-Planck-Institut f\"ur Astronomie, K\"onigstuhl 17, 69117 Heidelberg, Germany \\
    $^4$ UPMC-CNRS, UMR7095, Institut d'Astrophysique de Paris, 75014, Paris, France \\
    $^5$ Department of Physics, Kavli Institute for Astrophysics and Space Research, Massachusetts Institute of Technology, Cambridge, MA 02139, USA \\
    $^6$ Department of Physics and Astronomy, Rutgers University, 136 Frelinghuysen Rd., Piscataway, NJ 08854 \\
    $^7$ Universit\"ats-Sternwarte M\"unchen, Scheinerstrasse 1, 81679 M\"unchen, Germany \\
    $^8$ Max-Planck-Insitut f\"ur Extraterrestrische Physik, Giessenbachstrasse, 85748 Garching, Germany. \\
}

\begin{document}
\date{Accepted ... Received ...}

\maketitle

\begin{abstract} 
The distribution of galaxy morphological types is a key test for
models of galaxy formation and evolution, providing strong constraints
on the relative contribution of different physical processes
responsible for the growth of the spheroidal components. In this
paper, we make use of a suite of semi-analytic models to study the
efficiency of galaxy mergers in disrupting galaxy discs and building
galaxy bulges. In particular, we compare standard prescriptions
usually adopted in semi-analytic models, with new prescriptions
proposed by Kannan et al., based on results from high-resolution
hydrodynamical simulations, and we show that these new implementations
reduce the efficiency of bulge formation through mergers. In addition,
we compare our model results with a variety of observational
measurements of the fraction of spheroid-dominated galaxies as a
function of stellar and halo mass, showing that the present
uncertainties in the data represent an important limitation to our
understanding of spheroid formation. Our results indicate that the
main tension between theoretical models and observations does not stem
from the survival of purely disc structures (i.e. bulgeless galaxies),
rather from the distribution of galaxies of different morphological
types, as a function of their stellar mass.
\end{abstract}


\section{Introduction}\label{sec:intro}

The morphological classification of galaxies constitutes one of the
earliest attempts to explain the different observational properties of
galaxies in terms of an evolutionary sequence \citep{Hubble1926}. The
distribution of galaxies in different morphological classes has been
shown to correlate strongly both with local environment \citep[see
  e.g.][]{Dressler80, Wilman09} and stellar mass \citep[see
  e.g.][]{Vulcani11, WilmanErwin12}. Therefore modern astrophysical
research focuses on the complex interplay of physical mechanisms
acting on the baryonic components and responsible for the
morphological transformation as a function of stellar mass, cosmic
epoch and environment \citep{Brennan15}. Different approaches to the
morphological classification of galaxy samples have been proposed,
moving from the visual classification of (relatively) small samples of
local galaxies, to the use of automatic pipelines applied to galaxy
surveys. All these techniques have their strengths and shortcomings:
typically the merit of the classification heavily depends on
photometric data quality and/or on the algorithm used (e.g. profile or
2D fitting) and on the galactic components assumed. The aim is to
distinguish between a centrally concentrated spheroidal structure
(also defined as a ``bulge'') and a disc-like component (which may
also host spiral arms). This oversimplified picture neglects the
heterogeneity of the bulge population and the role played by bar-like
structures in galaxy evolution \citep[see
  e.g.][]{KormendyKennicutt04}, nevertheless it is useful to define
the mass ratio between the bulge mass and the total mass of the galaxy
(the so called bulge-to-total ratio, $B/T$) as a primary discriminant
between bulge- and disc-dominated galaxies.

Galaxies which consist purely of a bulge (ellipticals) are not easy to
identify unambiguously in observations, even in the very local
Universe. They have kinematic properties varying from pure
pressure-supported systems to systems with partial rotational support
similar to lenticular galaxies, just without the defining outer disk
component \citep{Krajnovic08, Emsellem11}. In absence of kinematics,
an accurate multi-component decomposition requires well resolved
imaging at high signal to noise, for which a 2-component (bulge+disk)
model can be fit \citep[see e.g.][]{Simard11, Mendel14}, providing
parameters tied to the morphological information. Visual
classifications - confirming the absence of disk light and asymmetric
features such as bars and spiral arms - provides an alternative
approach to identify ellipticals. Nonetheless, ellipticals are always
difficult to distinguish from face-on lenticular galaxies, and visual
classification requires a huge effort for the characterization of
large samples. In both cases, with the average B/T increasing to high
mass \citep[see e.g.][]{Bluck14}, it becomes more difficult to
distinguish pure elliptical galaxies from the increasing population of
intermediate B/T (and still fairly concentrated) galaxies. Therefore,
depending upon the exact cuts used to select elliptical or (more
inclusively) early-type galaxies, a different sample of elliptical
galaxies may be selected.

In a recent paper, \citet{WilmanErwin12} presented a local galaxy
catalogue, based on a visual classification of galaxy morphology. This
catalogue (SDSSRC3) includes galaxies with $M_\star>10^{10.5} M_\odot$
and it is based upon the New York University Value Added Galaxy
Catalogue (NYU-VAGC, \citealt{Blanton05}) who matched the SDSS DR4
(Sloan Digital Sky Survey Data Release 4,
\citealt{Adelman-McCarthy06}) to the Third Reference Catalogue of
Bright Galaxies (RC3,
\citealt{deVaucouleurs91}). \citet{WilmanErwin12} used this sample to
study the fraction of galaxies of different morphological types as a
function of stellar and parent Dark Matter (DM) halo mass. For central
galaxies, the fraction of (visually classified) elliptical galaxies is
found to increase with stellar mass and parent DM mass (only for
central galaxies). They also found a relatively small fraction of
genuine elliptical galaxies at high stellar masses ($M_\star>10^{11.5}
M_\odot$), with a large contribution of S0s (which never have
$B/T>0.7$) and Spiral galaxies. One of the main differences between
this work and previous ones lies in the definition of the galaxy
sample (relatively nearby and bright - they appear in the RC3
catalogue) and in the use of visual classification to separate S0s
from Ellipticals. Their definition is thus both qualitatively and
quantitatively different from alternative approaches to define samples
of Early Types galaxies based on fitting light profiles and/or
automatic decomposition (see e.g. \citealt{HydeBernardi09} for the
full SDSS volume).

The work by \citet{WilmanErwin12} has been extended in
\citet{Wilman13} including a detailed comparison to predictions from a
suite of semi-analytic models (SAMs hereafter) of galaxy formation and
evolution implementing different prescriptions for bulge
formation. They showed that theoretical predictions are able to
reproduce the observed increase of the Ellipticals fraction $f_E$ with
stellar/halo mass. However, theoretical $f_E$ is {\it too high} with
respect to the estimates from the SDSSRC3, by factors of a few (2.6 to
4.2 for the models considered in that paper). Moreover, models are
also able to reproduce the fraction of passive galaxies and the
fraction of star forming disc galaxies (the separation between star
forming and passive galaxies being set at a specific star formation
rate $sSFR=10^{-11} {\rm yr}$). Therefore, models overproduce
Elliptical galaxies at the expense of passive S0s and Spiral
galaxies. \citet{Wilman13} suggested a reduced efficiency of bulge
formation in mergers, via efficient stripping of satellite galaxies
leading to reduced baryonic merger ratios, as a viable solution to
this problem.

Contrasting results have been presented recently: \citet{Porter14}
compared predictions of improved variants of the \citet{Somerville08}
model with the fraction of Early Type galaxies computed using the
\citet{HydeBernardi09} SDSS sample, and conclude that their models
tend to {\it underpredict} the fraction of Early type galaxies, when
bulges are produced only in mergers. In view of these contradicting
claims, it becomes important to understand the origin of the
differences between observational determinations of the Early
Type/Elliptical fractions.

The efficiency of binary galaxy mergers in destroying discs and
forming bulges has been revised, following numerical simulations
taking into account the role of the gas content \citep[see
  e.g.][]{Robertson06, Governato09, Hopkins09a}. Consensus has grown
that the presence of a gas component may lead to the survival of the
original disc even in 1:1 mergers. However, these studies were based
on idealized merger configurations and did not cover a statistical
sample of cosmological initial conditions. As a matter of fact,
\citet{Moster11} repeated similiar simulations including both
cosmological accretion and accretion from a reservoir of hot gas
associated with the parent dark matter halo (two ingredients that were
not considered in \citealt{Hopkins09a}) and they did not find any
clear dependence of burst efficiency on the progenitor's cold gas
fraction.

A complementary approach has been used in the framework of the
so-called Simulated Merger Tree Approach \citep{Moster14}: this method
is based on high-resolution hydrodynamical simulations of galaxy
mergers, with initial conditions (Dark Matter structures, orbital
parameters, galaxy properties) extracted from (lower resolution)
cosmological simulations coupled with SAMs. This technique thus
combines the potential of high resolution to explore the effect of
mergers on the mass distribution of the remnant galaxy, with
cosmologically motivated initial conditions. Using this method,
\citet{Kannan15} analyzed the channels of bulge growth in 19
simulations of binary galaxy mergers. In particular, they followed the
transfer of stellar and cold gas mass from the satellite to the
central galaxy, as well as the mass exchanges between different
components (disc, bulge, halo) of the central galaxy. They then try to
quantify the mass flows by providing fitting formulae that parametrize
the dependence of such mass exchanges on the merger mass ratio (both
baryonic and in Dark Matter), and the orbital parameters. Their
results show that the morphology of merger remnant depends mainly on
the baryonic mass ratio between the two merging galaxies but also on
the gas fraction of the main galaxy, while the orbital parameters
(i.e. the eccentricity of the orbit) play a minor role.

The aim of this work is to study the implications of the new fitting
formulae when implemented within models that can create large
statistical realizations of the galaxy population, and in particular
the predicted fraction of early types as function of stellar and halo
mass. This paper is organized as follows. In Section~\ref{sec:models},
we introduce the semi-analytic implementations we consider in our
analysis. We then compare the different distribution of morphological
types to different sets of observational data in
Section~\ref{sec:results}. Finally, we discuss our conclusions in
Section~\ref{sec:final}.

\section{Models}\label{sec:models}
\begin{figure*}
  \centerline{ \includegraphics[width=18cm]{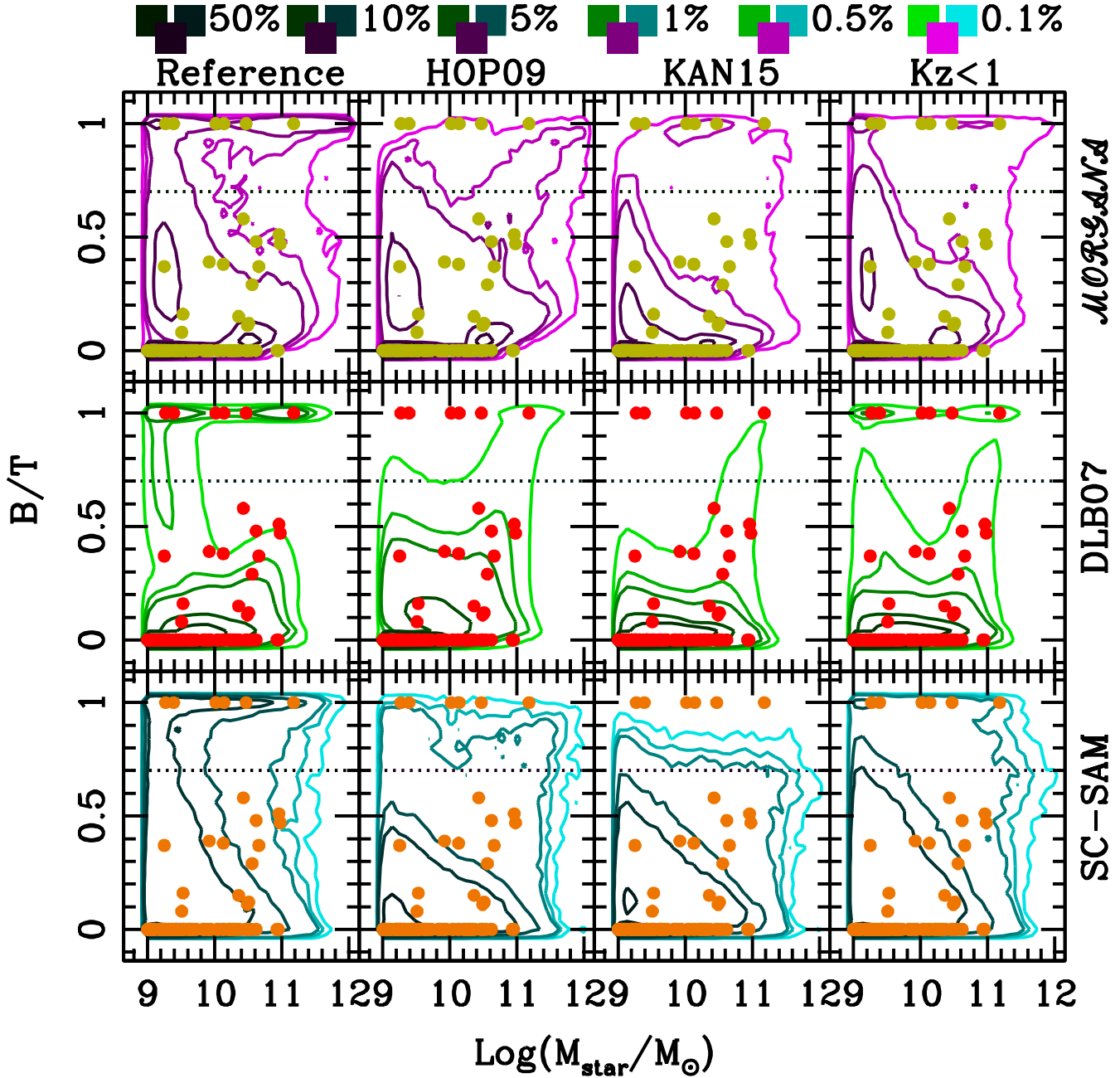} }
  \caption{Bimodal distribution of predicted galaxy morphologies as a
    function of stellar mass. Contour levels mark number density
    levels (normalized to the maximum density) as in the legend and
    the horizontal dotted line refers to $B/T=0.7$. Filled coloured
    circles show the ``classical'' bulge to total ratios of observed
    galaxies in the \citet{FisherDrory11} sample (i.e. we consider
    $B/T>0$ only for those bulges classified as ``classical'' in the
    original sample).}\label{fig:bimodal}
\end{figure*}

In this work we consider predictions from three different
state-of-the-art SAMs, namely \morgana \citep{Monaco07}, the
\citet[DLB07 hereafter]{DeLuciaBlaizot07} version of the \munich model
and the {\it Santa Cruz} model (\somer, \citealt{Somerville08,
  Porter14}). In particular, we consider the \morgana run defined in
\citet{LoFaro09} and calibrated on WMAP3 cosmology ($H_0 = 72$ km
s$^{-1}$ Mpc$^{-1}$ , $\Omega_m = 0.24$, $\sigma_8 = 0.8$, $n = 0.96$,
$\Omega_\Lambda = 0.76$); the DLB07 model applied to the Millennium
Simulation (\citealt{Springel05}, $H_0 = 73$ km s$^{-1}$ Mpc$^{-1}$ ,
$\Omega_m = 0.25$, $\sigma_8 = 0.9$, $n = 1$, $\Omega_\Lambda = 0.75$)
and the ``fiducial'' \somer model as presented in \citet[including
  also the modifications discussed in
  \citealt{Hirschmann12}]{Porter14} applied to the Bolshoi simulation
(\citealt{Klypin11}, $H_0 = 70.1$ km s$^{-1}$ Mpc$^{-1}$ , $\Omega_m =
0.279$, $\sigma_8 = 0.817$, $n = 0.96$, $\Omega_\Lambda = 0.721$). We
do not expect the differences in the cosmological parameters to affect
model predictions in a relevant way \cite[see e.g][]{Guo13,
  Wang08}. All models adopt a \citet{Chabrier03} IMF. We refer the
interested reader to the original papers for a complete overview of
the modelling of the different physical processes in each
code. Predictions from the same SAMs considered here have already been
compared against a number of observational constraints \citep[see
  e.g][]{Fontanot09b, Fontanot10}. In the following, we give a brief
description of the physical mechanisms relevant for bulge formation
and morphological transformations. A detailed analysis of the
timescales and channels of bulge formation in \morgana and DLB07 has
been presented in \citet{DeLucia11}, \citet{Fontanot11} and
\citet{Wilman13}.

\subsection{Reference and HOP09 runs}

All reference models assume that star formation happens primarily in
disc systems. Spheroidal systems form when the stars and/or gas in the
galactic discs lose angular momentum as a consequence of dissipative
processes like galaxy mergers and/or disc instabilities, and are
funnelled towards the centre of the galaxy. Broadly speaking, galaxy
mergers are supposed to lead to the formation of an Elliptical galaxy
or of a ``classical'' bulge (with kinematic and photometric properties
similar to those of Elliptical galaxies, see
e.g. \citealt{DaviesIllingworth83}), while secular processes are
associated with the formation of ``pseudo'' bulges (characterized by
disc-like exponential profiles or kinematics; see
e.g. \citealt{KormendyKennicutt04}).

The modelling of disc instabilities represents one of the largest
uncertainties in SAMs. Different assumptions are found in the
literature, from transferring just enough mass to restore stability
(e.g. \citealt{DeLuciaBlaizot07}) to the catastrophic destruction of
the entire disc into a spheroidal remnant \citep{Bower06}. Numerical
simulations of bar formation and evolution \citep[see
  e.g.][]{Debattista06} have not yet provided definitive evidence in
favour of a given treatment or alternative prescriptions (although
less dramatic mass transfers seem to be preferred Moster et al., in
preparation); given these uncertainties, in the following we switch
off disc instabilities in our reference models, that therefore
correspond to the pure merger realizations in \citet{Wilman13}. It
should be borne in mind that the contribution of disc instabilities is
negligible at the highest stellar masses, but relevant for
intermediate mass galaxies ($10^{10}<M_\star/M_\odot<10^{11}$) for all
models (as discussed in \citealt{DeLucia11} and \citealt{Porter14}),
thus playing a critical role for the interpretation of observational
data at these scales \citep{Fontanot11}. Moreover, in our models we
also do not consider tidal stripping of stars in satellite galaxies.

All 3 SAMs considered in this work implemented merger prescriptions
based on a broad division between major and minor mergers: in a major
merger (i.e. mass ratio $>$0.3) the entire stellar and gaseous content
of the two merging galaxies is given to the spheroidal remnant, which
becomes an Elliptical galaxy ($B/T=1$). In minor mergers, the stellar
mass of the satellite is given to the bulge of the central/remnant,
while the gaseous component is given either to the disc (DLB07,
\somer) or to the bulge (\morgana): in both cases the merger triggers
a starburst. \citet{Wilman13} showed that this scheme implies a
significant increase of $f_E$ with stellar mass and parent halo mass
for central galaxies. We refer to these model realizations as {\it
  reference} runs. In particular, for the SC-SAM reference run we use
the same merger prescriptions as in \citet{Somerville08}, which
includes the scattering of of a fixed ($0.2$) fraction of disc stars
into the diffuse stellar halo (DSH) at each galaxy merger.

A different formulation was proposed by \citet{Hopkins09a}, based on
results from idealized hydrodynamical simulations of binary mergers
\citep{Robertson06}. This ``gas-fraction-dependent merger model''
assumes that galaxy mergers trigger a burst of star formation, but the
fraction of cold gas that participates in the starburst
(i.e. efficiency of the conversion of cold gas into stars) depends on
both the baryonic merger ratio of the two progenitors, and on the gas
fraction of the disc of the primary galaxy. All stars formed in the
burst are assigned to the spheroidal component of the remnant galaxy,
as well as the whole stellar mass of the satellite galaxy. In
addition, the coalescence of the two galaxies might destroy a large
fraction of the primary disc (depending on the mass ratio of the
merger), transferring its mass to the spheroidal component of the
remnant. This is significantly different with respect to the reference
approach, where the disc of the primary galaxy is either unaffected in
minor mergers or completely destroyed in major
mergers. \citet{DeLucia11} studied the impact of the
\citet{Hopkins09a} prescriptions on bulge formation channels in the
framework of the \morgana and DLB07 models, finding small differences
in terms of galaxy distribution into different morphological types,
with respect to the reference runs. In contrast, \citet{Hopkins09b}
have investigated the impact of these prescriptions in the \somer
model, finding a suppression of bulge formation in low-mass galaxies,
and claiming that the gas dependence of merger-induced starbursts is a
fundamental ingredient to reproduce the observed morphology-mass
relation. An improved version of the \somer model including the
\citet{Hopkins09a} prescriptions has been presented in
\citet{Porter14}. We refer to this set of realizations as the HOP09
runs. While running the reference and KAN15/Kz$<$1 realizations for
the \somer model, we fixed all the other relevant parameters to the
values assumed in \citet{Porter14}.

The frequency of mergers is clearly related to the timescales assumed
for the orbital decay of satellites inside DM haloes. \morgana and
\somer assign the merger time at the last time the galaxy is central
(i.e. when the haloes merge), while in DLB07 the residual merging time
is estimated from the relative orbit of the two merging objects, at
the time of subhalo disruption. \morgana uses the fitting formula from
\citet{Taffoni03} while both DLB07 and \somer use variants of the
dynamical friction formula. In particular, \somer uses the formulation
proposed by \citet{BoylanKolchin08} and DLB07 use the Chandrasekhar
formula with a fudge factor 2 (which should bring merger times in
closer agreement with those predicted by the \citealt{BoylanKolchin08}
formula). The three approaches provide different predictions for the
dynamics of satellite galaxies \citep{DeLucia10}. In particular, the
merging timescale for massive satellites (i.e. with large progenitor
mass ratios) is roughly one order of magnitude shorter in \morgana
than in the other two models.

\subsection{Fitting formulae from \citet{Kannan15}.}
\begin{figure*}
  \centerline{ 
    \includegraphics[width=18cm]{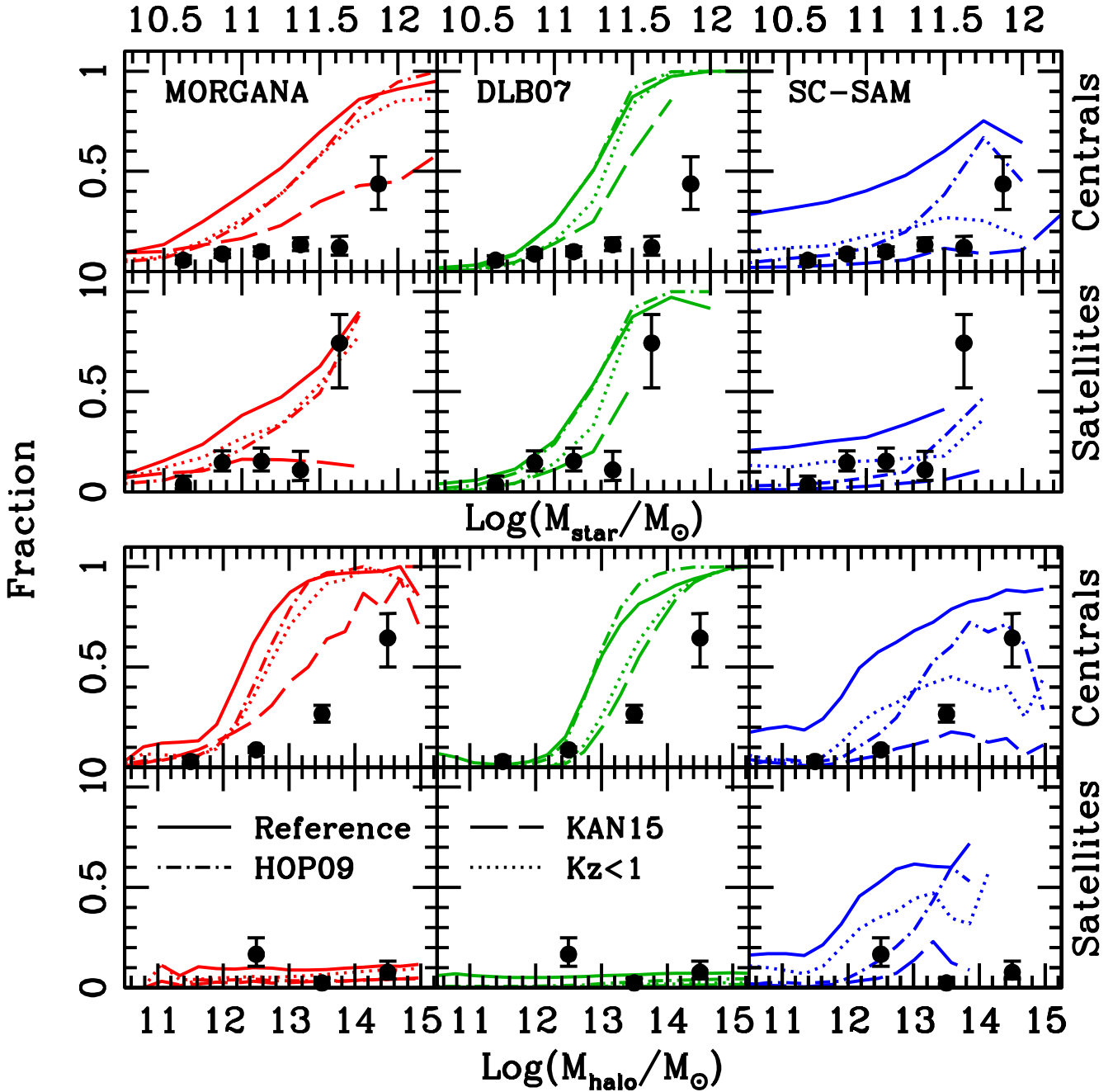} 
  }
  \caption{Elliptical galaxy fraction $f_E$ ($B/T>0.7$) as a function
    of stellar mass ({\it upper panels}) and parent halo mass ({\it
      lower panels}). Upper and lower rows refer to central and
    satellite galaxies respectively. In each panel, datapoints with
    errorbars correspond to visually classified total $f_E$ derived
    from the SDSSRC3 sample \citep{Wilman13}. Solid, dot-dashed,
    dashed and dotted lines refer to the SAMs predictions for the
    reference, HOP09, KAN15 and Kz$<$1 runs, respectively. Only bins
    with more than 10 model galaxies have been
    considered.}\label{fig:histograms}
\end{figure*}
\begin{figure*}
  \centerline{ 
    \includegraphics[width=18cm,height=9cm]{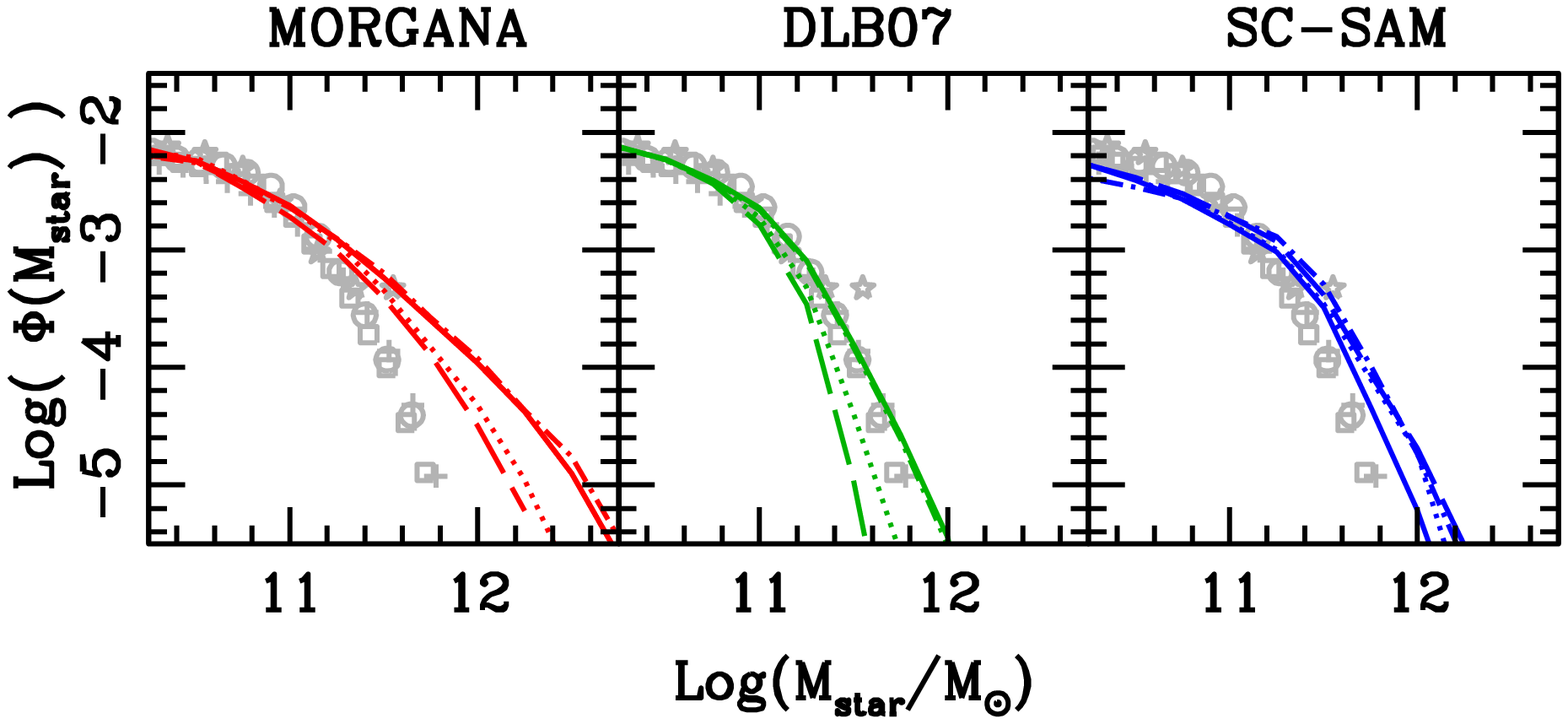} 
  }
  \caption{Galaxy stellar mass functions at $z=0$. Line types and
    colours refer to SAM predictions as in
    fig.~\ref{fig:histograms}. Light grey points refer to the stellar
    mass function compilation from \citet[see references
      herein]{Fontanot09b}.}\label{fig:mf}
\end{figure*}

We briefly recall here the main results from \citet{Kannan15}. Using
hydrodynamical simulations based on the parallel TreeSPH-code GADGET-2
\citep{Springel05}, they show that in binary galaxy mergers, the mass
exchanges between the satellite and the central galaxy and between the
different components of the central galaxy primarly depend on the
baryonic merger ratio ($\mu_{\rm b}$) between the two objects
(computed at the infall time of the satellite galaxy, i.e. when it
first became a satellite) and on the gas fraction ($f_{\rm gas}$) of
the primary galaxy. Following \citet{Moster14}, initial conditions for
the hydrodynamical simulations are set up at $z\sim1$ using \morgana
to populate DM merger trees constructed from a cosmological volume
using the Lagrangian code {\sc pinocchio} \citep{Monaco02}. \morgana
provides predictions for the physical properties of merging galaxies,
including stellar and cold gas mass, morphology and scale radii, while
infall times are derived directly from the DM merger tree. It is worth
stressing that the simulation suite used in \citet{Kannan15} analysis
includes only 19 merger configurations, extracted from the
cosmological volume and bound to have DM halo masses at $z=0$ similar
to the Milky Way host halo. Initial masses for primary galaxies lie in
the range $9.08<\log(M_\star)<10.77$, with merger ratios
$0.01<\mu_{\rm b}<1$ and gas fraction $0.1<f_{\rm gas}<0.6$.  This is
a rather small sample and results in a significant scatter around the
proposed analytic fitting formulae, thus limiting the precision of the
resulting fitting formulae. Moreover, since we will apply these
prescriptions on a full cosmological volume, we must extrapolate
beyond the DM halo mass scale and redshift range they were calibrated
on. Nonetheless, given the present uncertainties in the modeling of
bulge formation during galaxy mergers, it is interesting to explore
the implications of these trends in terms of the distribution of model
galaxies in the different morphological types, in order to assess if
these results go into the direction of relieving or exacerbating the
tension between data and model predictions. In order to test the
strength of our results, we define additional model realizations
imposing a (gaussian distributed) scatter around the adopted fitting
formulae corresponding to variance ranging from 0.2 to 1. The analysis
of these runs show that none of the conclusions we discuss in this
paper change when scatter in the fitting formulae is taken into
consideration.

In detail, the main mass exchanges parametrized in the
\citet{Kannan15} simulations include the fraction of total (stellar
plus cold gas) satellite mass ending up in the bulge of the remnant
galaxy

\begin{equation}
f_{\rm S2B} = \mu_{\rm b},
\end{equation}

\noindent
where $1-f_{\rm S2B}$ represents the satellite mass deposited into the
DSH as a result of tidal stripping acting on the secondary galaxy,
while it orbits within the parent DM halo potential well. They also
consider the fraction of the central disc gas mass transferred to the
central bulge

\begin{equation}
f_{\rm G2B} = (1-f_{\rm gas}) \mu_{\rm b},
\end{equation}

\noindent
which is strictly defined as $<1$. The cold gas transferred from the
primary disc to the bulge of the remnant, and the cold gas from the
satellite are readily consumed in a burst of star formation, following
the original definition in each model. Finally, the \citet{Kannan15}
fitting formulae describe the mass loss from the stellar disc of the
central galaxy, as the result of tidal effects: i.e. the fraction of
primary disc stars transferred to the bulge of the remnant

\begin{equation}
f_{\rm D2B} = 0.37 \mu_{\rm b},
\end{equation}

\noindent
and the fraction of central disc stars dispersed into the DSH

\begin{equation}
f_{\rm D2H} = 0.22 \mu_{\rm b}.
\end{equation}

\noindent
It is worth stressing that, for massive galaxies, results are more
sensitive to the fitting formulae describing the effect of mergers on
existing stars, as they typically have very low gas fractions.

In the following, we focus on two different sets of models: we first
consider SAMs realizations (KAN15) where the \citet{Kannan15} fitting
formulae are implemented at all redshifts. In a second set of modified
models (Kz$<$1), we allow these modifications only for $z<1$, keeping
the reference merger prescriptions at earlier times. This second
choice is motivated by the fact that the original simulations
considered in \citet{Kannan15} are set up at $z=1$ (using the
information provided by the SAM) and then run to $z=0$. Since we
expect the properties of higher-redshift discs to be different than
those of low-z systems, Kz$<$1 represents a conservative lower limit
on the effect of the \citet{Kannan15} prescriptions on the
distribution of morphological types.

\section{Results \& Discussion}\label{sec:results}

In fig.~\ref{fig:bimodal}, we compare the bidimensional $M_\star$ vs
$B/T$ distributions in our model galaxy catalogues and in the local
(closer than 11 Mpc) volume-limited sample of 321 nearby galaxies from
\citet{FisherDrory11}, based on a bulge-disc decomposition and
pseudo-bulge diagnostics in the {\it Spitzer} $3.6 \mu$m band. As disk
instability has been switched off in the models, we only consider
``classical'' $B/T$ ratios for the data sample (i.e. we consider
$B/T>0$ in the \citealt{FisherDrory11} sample only for those bulges
classified as classical, see also \citealt{Wilman13}). A consistent
picture can be drawn from the predictions of the different reference
models, despite different merger prescriptions and merger timescales.
In the KAN15 runs (i.e. when the new prescriptions are applied at all
redshifts) the abundance of $B/T>0.7$ galaxies is drastically reduced
at all mass scales. However, this is obtained at the expense of almost
completely devoiding the $B/T>0.9$ region. This effect goes into the
same direction as for the HOP09 runs, but the overall reduction of
bulge dominated galaxies is greater when using the \citet{Kannan15}
prescriptions in these models: the KAN15 column of
Fig.~\ref{fig:bimodal} shows that most model galaxies have
$B/T<0.5$. As expected, the Kz$<$1 model provides intermediate results
between the reference and KAN15 runs: the number of $B/T>0.7$ galaxies
is reduced, but not as much as in the KAN15 runs. In particular the
effect is mass dependent, as more massive galaxies are scattered to a
wider distribution of $B/T$ with respect to the reference runs (where
they are mostly $B/T=1$ objects).

More insight can be obtained by comparing the predictions of our
models with the observed $f_E$ as a function of stellar and parent DM
halo mass, as defined in \citet{Wilman13}. As in their work, in
Fig.~\ref{fig:histograms}, we define as Ellipticals all model galaxies
with $B/T>0.7$, and split our sample by hierarchy considering
separately central and satellite galaxies in the upper and lower rows
(only bins containing more that 10 objects are considered). Our choice
for the threshold at $B/T\sim0.7$ is motivated by observational
studies \citep{Weinzirl09, Laurikainen10}) and by consistency with
previous studies. We want to stress that our results do not change
qualitatively if we assume a different $B/T$ threshold for the
definition of Elliptical galaxies (in the range 0.5-0.9). We remind
the reader that the \citet{WilmanErwin12} sample contains 854 galaxies
with $M_\star>10^{10.5} M_\odot$, whose morphologies are visually
classified. It is difficult to directly compare this sample with the
\citet{FisherDrory11} catalogue, as the latter contains only 11
galaxies with $M_\star>10^{10.5} M_\odot$, while the former does not
allow for a finer classification between pseudo and classical
bulges. It is thus worth stressing that in the following plots $f_E$
formally represents an upper limit for the fraction of Elliptical
galaxies, as it is not possible to discriminate the relative
contribution of disk instabilities and merger processes in the
formation of a spheroid, but only its main shape (but see
\citealt{Erwin15} for recent progress on these issues). Nonetheless, a
more detailed classification should have a limited impact on our
conclusions, since we do not expect the formation of $B/T>0.7$ objects
to be driven by disk instabilities alone, although they might
contribute to increase the mass of a merger-induced spheroid. Similar
arguments hold for the other two main samples considered in this work
\citep{HydeBernardi09, Mendel14}.  As discussed in \citet{Wilman13},
the reference models (solid lines) overpredict $f_E$ over the whole
mass range. The discrepancy is larger for more massive systems, where
model fractions almost reach 100$\%$. The discrepancy is particularly
severe for central galaxies that show, in all models, a strong
dependence of $f_E$ on stellar and parent halo mass. It is interesting
to notice that the reference run of \somer (that was not included in
\citealt{Wilman13}) predicts lower elliptical fractions at the high
mass end, with respect to the other two models, thus providing the
closest match to the \citet{Wilman13} sample among reference
runs. Moreover, the predicted fractions of $B/T>0.7$ galaxies in the
HOP09 runs (dot-dashed lines) deviate considerably from the reference
runs only for the \somer model, while they are very close for the
other two models (as already noticed in \citealt{Wilman13}).

The picture changes when the KAN15 runs are considered: the predicted
$f_E$ are reduced at all mass scales, both as a function of stellar
and parent halo mass. It is worth stressing that the final masses of
model galaxies change between the reference and the KAN15/Kz$<$1 runs
(dashed and dotted lines), due to the different assembly history.  The
effect of these changes is imprinted on the $z=0$ galaxy stellar mass
function (Fig.~\ref{fig:mf}). For both \morgana and DLB07 there is a
clear decrease in stellar mass at the high-mass end due to the deposit
of stellar mass into the DSH as a consequence of the assumed modeling
for $f_{S2B}$ and $f_{D2H}$ (and because more massive galaxies
experience more merger events). This effect is more severe for DLB07,
as the number density of massive galaxies is already the smallest
among the models considered and the original prescription of this
model did not include stellar stripping of satellites (and we did not
retune the parameters to account for this effect). In this model, both
the KAN15 and Kz$<$1 runs do not predict any galaxy with stellar mass
$M_\star > 10^{11.5} M_\odot$, and the highest mass bins available are
affected by small number statistics. In addition, this model shows the
smallest reduction of $f_E$, with the $M_\star > 10^{11} M_\odot$ bins
still dominated by $B/T>0.7$ galaxies. We checked that, in DLB07, this
behaviour is mainly due to the reduced mass of infalling satellites
(i.e. $f_{S2B}$). We also noticed that the HOP09 runs of \morgana and
DBL07 predict galaxy stellar mass functions almost indistinguishable
from those of the reference runs. For the \somer instead, only small
differences between the predicted mass function from the four runs can
be seen. This is due to the fact that the \somer already includes the
scattering of disc stars into the DSH at each merger in the
\citet{Somerville08} and \citet{Porter14} versions. Therefore, the
impact of the \citet{Kannan15} formulae is reduced, with respect to
the other two models.

Despite for \morgana and \somer the predicted fractions for the KAN15
runs are now roughly consistent with the \citet{WilmanErwin12}
observational constraints, this is obtained at the expenses of losing
all $B/T>0.9$ Ellipticals.  The Kz$<$1 runs provide intermediate
results with respect to the other two sets of runs. In most cases,
however, the predicted $f_E$ still overpredicts the observed $f_E$ in
the \citet{WilmanErwin12} sample, at most mass scales.
\begin{figure*}
  \centerline{ 
    \includegraphics[width=18cm]{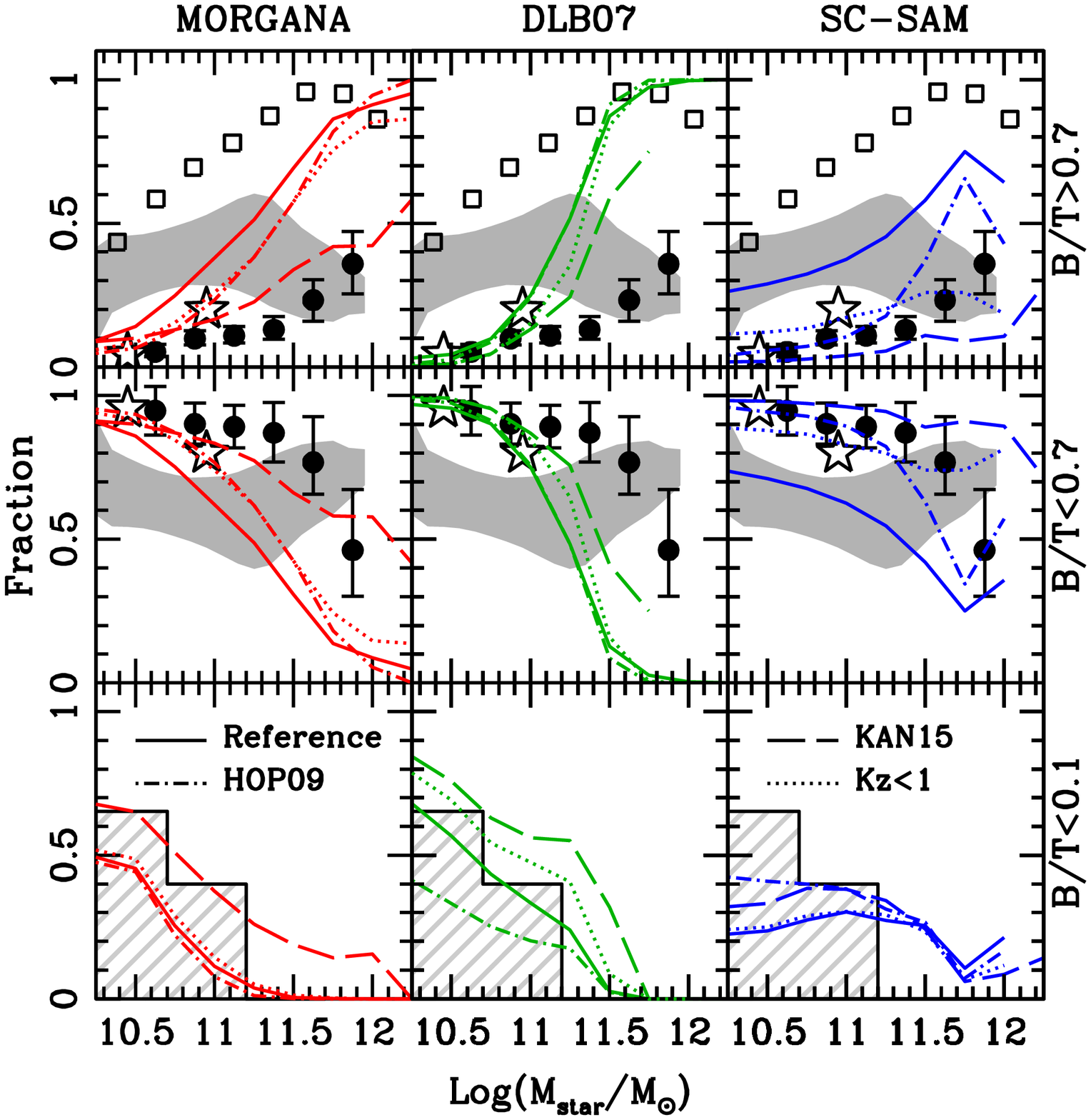} 
  }
  \caption{Distribution of galaxies among morphological types as a
    function of stellar mass. Filled circles refer to the total
    fraction of Ellipticals in \citet{Wilman13}, while open squares to
    the total fraction of Early Type galaxies from the sample of
    \citet{HydeBernardi09}. Shaded areas represent the results from
    the \citet{Mendel14} decomposition, while stars and hatched
    histograms refer to \citet{FisherDrory11} sample. See main text
    for more details on the observational samples considered. Line
    types and colours refer to SAM predictions as in
    fig.~\ref{fig:histograms}. Right-hand labels indicate the model
    galaxy $B/T$ range considered in each row.}\label{fig:total}
\end{figure*}

We then compare our model predictions with additional observational
constraints in Fig.~\ref{fig:total}. We first consider different $B/T$
ranges for model galaxies: we split the sample into $B/T>0.7$ (upper
row) and $B/T<0.7$ objects (mid row), and we also consider a $B/T<0.1$
selection (lower row), which corresponds to the fraction of
``bulgeless'' galaxies, i.e. those objects with the smallest
contribution from a spheroidal component. As for the observational
data, we consider four different samples and we plot their results in
each panel as follows. We first consider the \citet{WilmanErwin12}
sample, and we plot as solid circles their total $f_E$ in the upper
row. In the mid row, we plot the sum of the (visually classified) S$0$
and Spiral fractions. We include as open squares in the upper row, the
total fraction of Early Type galaxies in the SDSS selected by
\citet{HydeBernardi09} on the basis of the SDSS parameter fracDev=1
(the fraction of source light well-fitted by a de Vaucouleur's
profile) and $r$-band axis ratio $b/a>0.6$.

We also consider the catalogue of bulge, disk and total stellar mass
estimates for galaxies in the Legacy area of the SDSS presented in
\citet{Mendel14}. Those mass estimates are based on updated SDSS
photometry and bulge+disc decomposition, using dual S\'ersic profiles
\citep{Simard11}. In their analysis, \citet{Bluck14} consider a
subsample of the full catalogue defined by the cuts $0.02<z<0.2$,
$8<\log(M_{\rm star})<12$, $dDB<1$ and $P_{\rm pS}<0.5$. $dDB$
represents the scatter in the relation between the estimate of the
total stellar mass and the sum of the estimates of bulge and disc
separately; the applied cut removes objects with problematic
decomposition. $P_{\rm pS}$ is the probability that a source is
best-fit by a single S\'ersic profile; the applied cut removes sources
with possibly spurious components. It is worth stressing that the
bulge+disk modeling adopted in \citet{Mendel14}, i.e. the combination
of a de Vaucouleur's bulge and an exponential disk profiles, is unable
to accurately recover the total flux of galaxies with S\'ersic index
larger than 5. For these sources, the model fails in reproducing the
bright inner and extended outer profiles, at the same time.
Nonetheless, we include also these objects in our analysis, as they
might represent some extreme configurations of bulge/disc-dominated
sources. In order to assess for the uncertainties in their bulge+disc
decomposition, we define two subsamples of the \citet{Mendel14}
catalogues as follows. We first consider all galaxies satisfying the
first 3 cuts as in \citet{Bluck14}. We use this catalogue to define a
subsample (A) imposing $B/T=0$ to all objects with $P_{\rm pS} \ge
0.5$, in order to obtain a lower limit on the fraction of
bulge-dominated galaxies in the \citet{Mendel14} sample. Analogously,
we define a subsample (B) assuming $B/T=1$ for sources with $P_{\rm
  pS} \ge 0.5$, which maximizes the fraction of bulge-dominated
galaxies. The shaded area in Fig.~\ref{fig:total} represents the
confidence region between the resulting fractions of $B/T>0.7$ and
$B/T<0.7$ objects in the two subsamples. Fractions referring to
subsample A correspond to the lower and upper envelope of the shaded
region in the upper and mid panels respectively (viceversa for
subsample B).

Finally, we include results from \citet{FisherDrory11}. This local
sample allows a finer classification of spheroids into ``classical''
and ``pseudo''-bulges, thus providing a finer comparison with our
model predictions with respect to the \citet{Mendel14} sample, but it
contains only 11 galaxies in the stellar mass range of interest
(i.e. $M_\star>10^{11.5} M_\odot$). As in all our runs we switched off
disc instabilities, our models formally account for the formation of
classical bulges only, i.e. we assume that the entire ``pseudo''-bulge
population is the result of secular processes. We first plot the
resulting morphological fractions as stars in the upper and middle
panels of Fig.~\ref{fig:total}. These fractions do not depend on the
bulge classification, since there are no galaxies hosting
pseudo-bulges with $B/T>0.7$, and in reasonable agreement with both
\citet{WilmanErwin12} and \citet{Mendel14}. The hatched histograms in
the lower row represents (as in \citealt{Fontanot11}) the fraction of
galaxies with classical $B/T<0.1$ in the \citet{FisherDrory11}
sample. We stress that the sample contains a significant number of
galaxies hosting pseudo-bulges with $B/T>0.3$, which may hide a
(subdominant) merger-driven classical bulge. This effect would tend to
reduce the estimated fraction of bulgeless galaxies and the hatched
histograms then represent an upper limit to the bulgeless fraction in
the sample.

This comparison is meant to account for the uncertainty in the
definition of galaxy morphology. In general, there are still large
differences in the observational estimates, which explain the
apparently contradictory conclusions given in~\citet{Porter14}
and~\citet{Wilman13}. The tension between different samples is likely
due to the different techniques adopted to estimate the relative
importance of bulges and discs. The biggest discrepancy is seen in the
upper row between the Early Type catalogue of \citet{HydeBernardi09}
and the Elliptical fraction from the \citet{WilmanErwin12} sample; the
\citet{Mendel14} decomposition shows intermediate results, but is in
better agreement with the latter sample at the high mass end. In the
mid row, we show how the reduced bulge formation efficiency in galaxy
mergers impacts the statistics of disc-dominated galaxies. As in
Fig.~\ref{fig:histograms}, we see that the KAN15 run of the \morgana
model and the \somer runs provide the best agreement with the
\citet{WilmanErwin12} sample, also when $B/T<0.7$ model galaxies are
compared with sources with a relevant disc component (either S$0$s or
spiral galaxies). However, none of the models and runs considered in
this work is able to reproduce the peculiar trends of the
\citet{Mendel14} sample.

Finally, in the lower row, we focus on the number density of
``bulgeless'' galaxies. The formation of these objects has been
claimed to represent a potential challenge for the standard
cosmological model \citep[see e.g.][and references
  herein]{Kormendy10}. We showed in \citet{Fontanot11} that these
objects do not constitute a rare population in SAMs, as there is a
sufficient number of DM halos with merger-quiet assembly histories to
account for local statistics \citep[see also][]{Porter14}. With
respect to the reference runs, both the KAN15 and Kz$<$1 realizations
predict an increase of the fraction of $B/T<0.1$ galaxies at all mass
scales, which improves the agreement with the observations. The
increase is relatively mild in DLB07 and \somer and more marked in
\morgana. This is due to the shorter timescales for merging assumed by
the latter model, which induce more galaxy mergers.

\begin{figure}
  \centerline{ 
    \includegraphics[width=9cm]{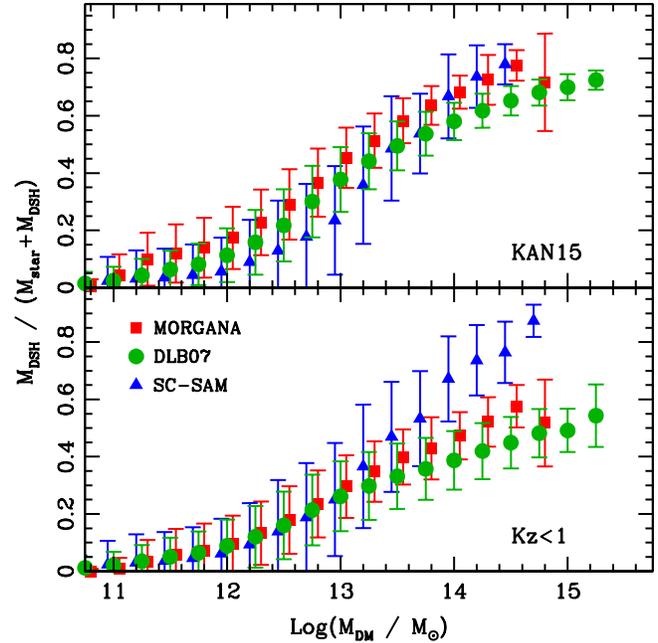} 
  }
  \caption{Fraction of stellar mass scattered into the DSH component
    (relative to the mass of the central galaxy plus the DSH) as a
    function of the mass of the parent DM halo.}\label{fig:dsh}
\end{figure}

A major role in the \citet{Kannan15} formulation is played by the
stars getting stripped (from the satellite galaxies) or scattered
(from the primary discs), and dispersed into the parent DM halo. In
Fig.~\ref{fig:dsh}, we show the stellar mass in the DSH as a function
of halo mass, normalized to the stellar mass of the central galaxy
plus the DSH itself. The upper panel refers to KAN15 runs, while the
lower panel to Kz$<$1 runs. In all cases, there is a clear trend of a
larger contribution of DSH at increasing halo masses. A direct
comparison of these DSH estimates with observational constraints is
complicated by the intrinsic difficulties in disentangling the
contribution to the total cluster luminosities from the central bright
galaxy and from the DSH (i.e. the intra cluster light), due to the
different techniques and assumptions \citep[see e.g.][for a review on
  these issues]{Zibetti08}. At face value, our predictions agree well
with constraints from \citet[$\gtrsim 0.6$ at $M_{\rm DM} \gtrsim
  10^{14} M_\odot$]{Gonzalez05}, but are larger than the recent
estimates from \citet[$\sim 0.2-0.4$]{Budzynski14}. Moreover, the
differences between the KAN15 and Kz$<$1 runs are small, showing that
in these models, most of the DSH is formed at $z<1$. This confirms
that most of the differences at the high mass end between the
reference models and the new runs are due to the effect of stellar
stripping/dispersion, which reduce the mass of both progenitors,
rather than to the influence of gas fraction (as noted above, most
model massive galaxies have typically low gas fractions at $z<1$).

\section{Conclusions}
\label{sec:final}

In this paper we study the impact of a different modelling of baryonic
(stellar and cold gas) mass transfers associated with galaxy mergers
on the distribution of morphological types as a function of hierarchy,
stellar mass and parent DM mass. We consider three independently
developed SAMs, namely the \somer model, the DLB07 version of the
\munich model and \morgana. When compared to a local sample of
visually classified galaxies, the two latter models have already been
shown to overpredict the fraction of Elliptical galaxies at high
masses, while reproducing the observed fraction of passive
galaxies. \citet{Wilman13} ascribed this discrepancy to a too high
efficiency of morphological transformation (from disc- to
bulge-dominated system) during galaxy mergers (which they suggest can
be reduced by assuming that a fraction of the satellite galaxy is
deposited into the DSH before the merger takes place). On the other
hand, \citet{Porter14} showed that the \somer model underpredicts the
fraction of Early Type galaxies in the \citet{HydeBernardi09} SDSS
sample. In this work, we extend the analysis by comparing model
predictions to four different observational samples
\citep{HydeBernardi09, FisherDrory11, WilmanErwin12, Mendel14} and by
considering new prescriptions for the mass exchanges during galaxy
merges, recently proposed by \citet{Kannan15}. These authors analyzed
high-resolution hydrodynamical simulations of $z<1$ binary mergers,
with initial condition extracted from a combination of cosmological
simulations and SAMs. The simulations were used to parametrize the
strength of baryonic mass transfer between the different components
(disc, bulge, halo) as a function of an effective merger ratio (taking
into account both the intrinsic merger ratio and the details of the
relative orbit).

Our results confirm that the different modelling of mass transfer
during mergers leads to significant modification of the distribution
of $B/T$ ratios as a function of both stellar and parent halo mass. In
particular, the fraction of $B/T>0.7$ galaxies, $f_E$, is reduced at
all mass scales. However, runs using the \citet{Kannan15} fitting
formulae at all redshifts result in virtually no galaxy with
$B/T=1$. Moreover, in these model realizations most of the bulge mass
is assembled at $z>1$. More conservative runs, enabling the new
fitting formulae only at $z<1$ (i.e. the redshift range probed by the
simulations) provide a better match to the $B/T$ distribution, at the
expense of a smaller reduction of $f_E$. These results suggest the
need to extend the \citet{Kannan15} work at higher redshifts. In fact,
$z>1$ discs are likely very different structures with respect to local
Spirals \citep[see e.g.][]{ForsterSchreiber09}, therefore we expect
them to show a different response to the tidal fields triggered by
close encounters, with respect to their local counterparts.

The detailed effects of the new prescriptions on the distribution of
galaxy morphologies are strongly SAM dependent. \morgana shows a
significant reduction of $f_E$ which moves the model predictions in
better agreement with the \citet{Wilman13} data. In DLB07, the most
relevant effect is that of decreasing the mass of the most massive
galaxies, due to the stripping effects included in \citet{Kannan15}
formulae. At intermediate-to-low mass scales, model predictions are
similar to the SDSSRC3 fractions, while at higher masses they are
closer to the early type fraction in \citet{HydeBernardi09} than to
the $f_E$ estimate from SDSSRC3. Finally, the effect of the new
prescriptions is limited in \somer due to the fact that the reference
run already predicts a reduced $f_E$ with respect to the other two
SAMs we consider: this model still undepredicts the Early Type
fractions as in \citet{HydeBernardi09}, but its predictions are
similar to the $f_E$ values from SDSSRC3.

Overall, the new formulation predicts a decrease of the fraction of
bulge-dominated ($B/T>0.7$) galaxies and a corresponding increase in
the fraction of disc-dominated galaxies (including bulgeless
galaxies). The comparison between model predictions and data is
complicated by the different observational estimates for the fraction
of bulge-dominated objects. As an additional test, in this work we
also consider the morphological fractions coming from the
\citet{Mendel14} sample. In general, this sample provides
morphological fractions which are intermediate with respect to the
\citet{WilmanErwin12} and \citet{HydeBernardi09} datasets, but at the
high mass end, bulge+disk decomposition in this sample is consistent
with the results based on visual inspection of SDSSRC3 galaxies. It is
worth stressing that the spread between the different observational
constraints for $f_E$ is significant. Much of this tension is directly
linked to the different techniques used to estimate galaxy morphology,
and ranging from purely automatic classification to human visual
inspection. Additional sources of uncertainty come from the presence
of fine structure features like spiral arms and bars, whose detection
strongly depend on the data quality the classification is performed on
(independently from the overall strategy). Much work is therefore
needed, on the observational side, in order to assess the consistency
of the different classifications and/or to understand the systematics
involved in each choice. The end product of this process will provide
even stronger constraints for modelers trying to quantify the amount
of morphological transformation connected to the different channels of
bulge growth.

We stress that the \citet{Kannan15} fitting formulae are based on a
small sample of high-resolution hydrodynamical simulations (19 binary
isolated mergers, covering a fair range of merger ratios), and their
results show a large scatter around the fitted relations. More
high-resolution simulations are thus required to better quantify both
the mean relations and their scatter, which could also play a role in
defining the observed distribution of morphological types. Moreover,
galaxy mergers in SAMs are assumed to be binary, i.e. involving only
two galaxies. This is an idealized assumption useful for most physical
configurations, but not representative when more than one satellite
interacts with the central galaxy at the same time \citep[see
  e.g.][]{Villalobos14}. More work in this direction is needed to
characterize mass transfers in multiple merger configurations, as a
function of the relative masses and orbits.

\section*{Acknowledgements}

FF acknowledges financial support from the grants PRIN MIUR 2009 ``The
Intergalactic Medium as a probe of the growth of cosmic structures'',
PRIN INAF 2010 ``From the dawn of galaxy formation'' and from the
Klaus Tschira Foundation. MH acknowledges financial support from the
European Research Council via an Advanced Grant under grant agreement
no. 321323 (NEOGAL). RSS thanks the Downsbrough family for their
generous support. DJW thanks the Deutsche Forschungsgemeinschaft for
funding via project WI 3871/1-1

\bibliographystyle{mn2e}
\bibliography{fontanot}

\begin{thebibliography}{}

\bibitem[\protect\citeauthoryear{{Adelman-McCarthy}, {Ag{\"u}eros}, {Allam},
  {Anderson}, {Anderson}, {Annis}, {Bahcall} \& {Baldry}}{{Adelman-McCarthy}
  et~al.}{2006}]{Adelman-McCarthy06}
{Adelman-McCarthy} J.~K.,  {Ag{\"u}eros} M.~A.,  {Allam} S.~S.,  {Anderson}
  K.~S.~J.,  {Anderson} S.~F.,  {Annis} J.,  {Bahcall} N.~A.,    {Baldry} I.~K.
  e.~a.,  2006, \apjs, 162, 38

\bibitem[\protect\citeauthoryear{{Blanton}, {Schlegel}, {Strauss}, {Brinkmann},
  {Finkbeiner}, {Fukugita}, {Gunn}, {Hogg}, {Ivezi{\'c}}, {Knapp}, {Lupton},
  {Munn}, {Schneider}, {Tegmark} \& {Zehavi}}{{Blanton}
  et~al.}{2005}]{Blanton05}
{Blanton} M.~R.,  {Schlegel} D.~J.,  {Strauss} M.~A.,  {Brinkmann} J.,
  {Finkbeiner} D.,  {Fukugita} M.,  {Gunn} J.~E.,  {Hogg} D.~W.,  {Ivezi{\'c}}
  {\v Z}.,  {Knapp} G.~R.,  {Lupton} R.~H.,  {Munn} J.~A.,  {Schneider} D.~P.,
  {Tegmark} M.,    {Zehavi} I.,  2005, \aj, 129, 2562

\bibitem[\protect\citeauthoryear{{Bluck}, {Mendel}, {Ellison}, {Moreno},
  {Simard}, {Patton} \& {Starkenburg}}{{Bluck} et~al.}{2014}]{Bluck14}
{Bluck} A.~F.~L.,  {Mendel} J.~T.,  {Ellison} S.~L.,  {Moreno} J.,  {Simard}
  L.,  {Patton} D.~R.,    {Starkenburg} E.,  2014, \mnras, 441, 599

\bibitem[\protect\citeauthoryear{{Bower}, {Benson}, {Malbon}, {Helly}, {Frenk},
  {Baugh}, {Cole} \& {Lacey}}{{Bower} et~al.}{2006}]{Bower06}
{Bower} R.~G.,  {Benson} A.~J.,  {Malbon} R.,  {Helly} J.~C.,  {Frenk} C.~S.,
  {Baugh} C.~M.,  {Cole} S.,    {Lacey} C.~G.,  2006, \mnras, 370, 645

\bibitem[\protect\citeauthoryear{{Boylan-Kolchin}, {Ma} \&
  {Quataert}}{{Boylan-Kolchin} et~al.}{2008}]{BoylanKolchin08}
{Boylan-Kolchin} M.,  {Ma} C.-P.,    {Quataert} E.,  2008, \mnras, 383, 93

\bibitem[\protect\citeauthoryear{{Brennan}, {Pandya}, {Somerville}, {Barro},
  {Taylor}, {Wuyts}, {Bell}, {Dekel}, {Ferguson}, {McIntosh}, {Papovich} \&
  {Primack}}{{Brennan} et~al.}{2015}]{Brennan15}
{Brennan} R.,  {Pandya} V.,  {Somerville} R.~S.,  {Barro} G.,  {Taylor} E.~N.,
  {Wuyts} S.,  {Bell} E.~F.,  {Dekel} A.,  {Ferguson} H.~C.,  {McIntosh} D.~H.,
   {Papovich} C.,    {Primack} J.,  2015, ArXiv e-prints

\bibitem[\protect\citeauthoryear{{Budzynski}, {Koposov}, {McCarthy} \&
  {Belokurov}}{{Budzynski} et~al.}{2014}]{Budzynski14}
{Budzynski} J.~M.,  {Koposov} S.~E.,  {McCarthy} I.~G.,    {Belokurov} V.,
  2014, \mnras, 437, 1362

\bibitem[\protect\citeauthoryear{{Chabrier}}{{Chabrier}}{2003}]{Chabrier03}
{Chabrier} G.,  2003, \apjl, 586, L133

\bibitem[\protect\citeauthoryear{{Davies} \& {Illingworth}}{{Davies} \&
  {Illingworth}}{1983}]{DaviesIllingworth83}
{Davies} R.~L.,  {Illingworth} G.,  1983, \apj, 266, 516

\bibitem[\protect\citeauthoryear{{De Lucia} \& {Blaizot}}{{De Lucia} \&
  {Blaizot}}{2007}]{DeLuciaBlaizot07}
{De Lucia} G.,  {Blaizot} J.,  2007, \mnras, 375, 2

\bibitem[\protect\citeauthoryear{{De Lucia}, {Boylan-Kolchin}, {Benson},
  {Fontanot} \& {Monaco}}{{De Lucia} et~al.}{2010}]{DeLucia10}
{De Lucia} G.,  {Boylan-Kolchin} M.,  {Benson} A.~J.,  {Fontanot} F.,
  {Monaco} P.,  2010, \mnras, 406, 1533

\bibitem[\protect\citeauthoryear{{De Lucia}, {Fontanot}, {Wilman} \&
  {Monaco}}{{De Lucia} et~al.}{2011}]{DeLucia11}
{De Lucia} G.,  {Fontanot} F.,  {Wilman} D.,    {Monaco} P.,  2011, \mnras,
  414, 1439

\bibitem[\protect\citeauthoryear{{de Vaucouleurs}, {de Vaucouleurs}, {Corwin}
  Jr., {Buta}, {Paturel} \& {Fouque}}{{de Vaucouleurs}
  et~al.}{1991}]{deVaucouleurs91}
{de Vaucouleurs} G.,  {de Vaucouleurs} A.,  {Corwin} Jr. H.~G.,  {Buta} R.~J.,
  {Paturel} G.,    {Fouque} P.,  1991, \skytel, 82, 621

\bibitem[\protect\citeauthoryear{{Debattista}, {Mayer}, {Carollo}, {Moore},
  {Wadsley} \& {Quinn}}{{Debattista} et~al.}{2006}]{Debattista06}
{Debattista} V.~P.,  {Mayer} L.,  {Carollo} C.~M.,  {Moore} B.,  {Wadsley} J.,
    {Quinn} T.,  2006, \apj, 645, 209

\bibitem[\protect\citeauthoryear{{Dressler}}{{Dressler}}{1980}]{Dressler80}
{Dressler} A.,  1980, \apj, 236, 351

\bibitem[\protect\citeauthoryear{{Emsellem}, {Cappellari}, {Krajnovi{\'c}},
  {Alatalo}, {Blitz}, {Bois}, {Bournaud}, {Bureau} \& et al.}{{Emsellem}
  et~al.}{2011}]{Emsellem11}
{Emsellem} E.,  {Cappellari} M.,  {Krajnovi{\'c}} D.,  {Alatalo} K.,  {Blitz}
  L.,  {Bois} M.,  {Bournaud} F.,  {Bureau} M.,    et al. 2011, \mnras, 414,
  888

\bibitem[\protect\citeauthoryear{{Erwin}, {Saglia}, {Fabricius}, {Thomas},
  {Nowak}, {Rusli}, {Bender}, {Vega Beltr{\'a}n} \& {Beckman}}{{Erwin}
  et~al.}{2015}]{Erwin15}
{Erwin} P.,  {Saglia} R.~P.,  {Fabricius} M.,  {Thomas} J.,  {Nowak} N.,
  {Rusli} S.,  {Bender} R.,  {Vega Beltr{\'a}n} J.~C.,    {Beckman} J.~E.,
  2015, \mnras, 446, 4039

\bibitem[\protect\citeauthoryear{{Fisher} \& {Drory}}{{Fisher} \&
  {Drory}}{2011}]{FisherDrory11}
{Fisher} D.~B.,  {Drory} N.,  2011, \apjl, 733, L47+

\bibitem[\protect\citeauthoryear{{Fontanot}, {De Lucia}, {Monaco}, {Somerville}
  \& {Santini}}{{Fontanot} et~al.}{2009}]{Fontanot09b}
{Fontanot} F.,  {De Lucia} G.,  {Monaco} P.,  {Somerville} R.~S.,    {Santini}
  P.,  2009, \mnras, 397, 1776

\bibitem[\protect\citeauthoryear{{Fontanot}, {De Lucia}, {Wilman} \&
  {Monaco}}{{Fontanot} et~al.}{2011}]{Fontanot11}
{Fontanot} F.,  {De Lucia} G.,  {Wilman} D.,    {Monaco} P.,  2011, \mnras,
  416, 409

\bibitem[\protect\citeauthoryear{{Fontanot}, {Pasquali}, {De Lucia}, {van den
  Bosch}, {Somerville} \& {Kang}}{{Fontanot} et~al.}{2011}]{Fontanot10}
{Fontanot} F.,  {Pasquali} A.,  {De Lucia} G.,  {van den Bosch} F.~C.,
  {Somerville} R.~S.,    {Kang} X.,  2011, \mnras, 413, 957

\bibitem[\protect\citeauthoryear{{F{\"o}rster Schreiber}, {Genzel},
  {Bouch{\'e}}, {Cresci}, {Davies}, {Buschkamp}, {Shapiro} \&
  {Tacconi}}{{F{\"o}rster Schreiber} et~al.}{2009}]{ForsterSchreiber09}
{F{\"o}rster Schreiber} N.~M.,  {Genzel} R.,  {Bouch{\'e}} N.,  {Cresci} G.,
  {Davies} R.,  {Buschkamp} P.,  {Shapiro} K.,    {Tacconi} L.~J. e.~a.,  2009,
  \apj, 706, 1364

\bibitem[\protect\citeauthoryear{{Gonzalez}, {Zabludoff} \&
  {Zaritsky}}{{Gonzalez} et~al.}{2005}]{Gonzalez05}
{Gonzalez} A.~H.,  {Zabludoff} A.~I.,    {Zaritsky} D.,  2005, \apj, 618, 195

\bibitem[\protect\citeauthoryear{{Governato}, {Brook}, {Brooks}, {Mayer},
  {Willman}, {Jonsson}, {Stilp}, {Pope}, {Christensen}, {Wadsley} \&
  {Quinn}}{{Governato} et~al.}{2009}]{Governato09}
{Governato} F.,  {Brook} C.~B.,  {Brooks} A.~M.,  {Mayer} L.,  {Willman} B.,
  {Jonsson} P.,  {Stilp} A.~M.,  {Pope} L.,  {Christensen} C.,  {Wadsley} J.,
   {Quinn} T.,  2009, \mnras, 398, 312

\bibitem[\protect\citeauthoryear{{Guo}, {White}, {Angulo}, {Henriques},
  {Lemson}, {Boylan-Kolchin}, {Thomas} \& {Short}}{{Guo} et~al.}{2013}]{Guo13}
{Guo} Q.,  {White} S.,  {Angulo} R.~E.,  {Henriques} B.,  {Lemson} G.,
  {Boylan-Kolchin} M.,  {Thomas} P.,    {Short} C.,  2013, \mnras, 428, 1351

\bibitem[\protect\citeauthoryear{{Hirschmann}, {Somerville}, {Naab} \&
  {Burkert}}{{Hirschmann} et~al.}{2012}]{Hirschmann12}
{Hirschmann} M.,  {Somerville} R.~S.,  {Naab} T.,    {Burkert} A.,  2012,
  \mnras, 426, 237

\bibitem[\protect\citeauthoryear{{Hopkins}, {Cox}, {Younger} \&
  {Hernquist}}{{Hopkins} et~al.}{2009}]{Hopkins09a}
{Hopkins} P.~F.,  {Cox} T.~J.,  {Younger} J.~D.,    {Hernquist} L.,  2009,
  \apj, 691, 1168

\bibitem[\protect\citeauthoryear{{Hopkins}, {Somerville}, {Cox}, {Hernquist},
  {Jogee}, {Kere{\v s}}, {Ma}, {Robertson} \& {Stewart}}{{Hopkins}
  et~al.}{2009}]{Hopkins09b}
{Hopkins} P.~F.,  {Somerville} R.~S.,  {Cox} T.~J.,  {Hernquist} L.,  {Jogee}
  S.,  {Kere{\v s}} D.,  {Ma} C.,  {Robertson} B.,    {Stewart} K.,  2009,
  \mnras, 397, 802

\bibitem[\protect\citeauthoryear{{Hubble}}{{Hubble}}{1926}]{Hubble1926}
{Hubble} E.~P.,  1926, \apj, 64, 321

\bibitem[\protect\citeauthoryear{{Hyde} \& {Bernardi}}{{Hyde} \&
  {Bernardi}}{2009}]{HydeBernardi09}
{Hyde} J.~B.,  {Bernardi} M.,  2009, \mnras, 394, 1978

\bibitem[\protect\citeauthoryear{{Kannan}, {Macci\`o}, {Fontanot}, {Moster},
  {Karman} \& {Somerville}}{{Kannan} et~al.}{2015}]{Kannan15}
{Kannan} R.,  {Macci\`o} A.,  {Fontanot} F.,  {Moster} B.,  {Karman} W.,
  {Somerville} R.,  2015, MNRAS submitted

\bibitem[\protect\citeauthoryear{{Klypin}, {Trujillo-Gomez} \&
  {Primack}}{{Klypin} et~al.}{2011}]{Klypin11}
{Klypin} A.~A.,  {Trujillo-Gomez} S.,    {Primack} J.,  2011, \apj, 740, 102

\bibitem[\protect\citeauthoryear{{Kormendy}, {Drory}, {Bender} \&
  {Cornell}}{{Kormendy} et~al.}{2010}]{Kormendy10}
{Kormendy} J.,  {Drory} N.,  {Bender} R.,    {Cornell} M.~E.,  2010, \apj, 723,
  54

\bibitem[\protect\citeauthoryear{{Kormendy} \& {Kennicutt} Jr.}{{Kormendy} \&
  {Kennicutt}}{2004}]{KormendyKennicutt04}
{Kormendy} J.,  {Kennicutt} Jr. R.~C.,  2004, \araa, 42, 603

\bibitem[\protect\citeauthoryear{{Krajnovi{\'c}}, {Bacon}, {Cappellari},
  {Davies}, {de Zeeuw}, {Emsellem}, {Falc{\'o}n-Barroso}, {Kuntschner},
  {McDermid}, {Peletier}, {Sarzi}, {van den Bosch} \& {van de
  Ven}}{{Krajnovi{\'c}} et~al.}{2008}]{Krajnovic08}
{Krajnovi{\'c}} D.,  {Bacon} R.,  {Cappellari} M.,  {Davies} R.~L.,  {de Zeeuw}
  P.~T.,  {Emsellem} E.,  {Falc{\'o}n-Barroso} J.,  {Kuntschner} H.,
  {McDermid} R.~M.,  {Peletier} R.~F.,  {Sarzi} M.,  {van den Bosch} R.~C.~E.,
    {van de Ven} G.,  2008, \mnras, 390, 93

\bibitem[\protect\citeauthoryear{{Laurikainen}, {Salo}, {Buta}, {Knapen} \&
  {Comer{\'o}n}}{{Laurikainen} et~al.}{2010}]{Laurikainen10}
{Laurikainen} E.,  {Salo} H.,  {Buta} R.,  {Knapen} J.~H.,    {Comer{\'o}n} S.,
   2010, \mnras, 405, 1089

\bibitem[\protect\citeauthoryear{{Lo Faro}, {Monaco}, {Vanzella}, {Fontanot},
  {Silva} \& {Cristiani}}{{Lo Faro} et~al.}{2009}]{LoFaro09}
{Lo Faro} B.,  {Monaco} P.,  {Vanzella} E.,  {Fontanot} F.,  {Silva} L.,
  {Cristiani} S.,  2009, \mnras, 399, 827

\bibitem[\protect\citeauthoryear{{Mendel}, {Simard}, {Palmer}, {Ellison} \&
  {Patton}}{{Mendel} et~al.}{2014}]{Mendel14}
{Mendel} J.~T.,  {Simard} L.,  {Palmer} M.,  {Ellison} S.~L.,    {Patton}
  D.~R.,  2014, \apjs, 210, 3

\bibitem[\protect\citeauthoryear{{Monaco}, {Fontanot} \& {Taffoni}}{{Monaco}
  et~al.}{2007}]{Monaco07}
{Monaco} P.,  {Fontanot} F.,    {Taffoni} G.,  2007, \mnras, 375, 1189

\bibitem[\protect\citeauthoryear{{Monaco}, {Theuns}, {Taffoni}, {Governato},
  {Quinn} \& {Stadel}}{{Monaco} et~al.}{2002}]{Monaco02}
{Monaco} P.,  {Theuns} T.,  {Taffoni} G.,  {Governato} F.,  {Quinn} T.,
  {Stadel} J.,  2002, \apj, 564, 8

\bibitem[\protect\citeauthoryear{{Moster}, {Macci{\`o}} \&
  {Somerville}}{{Moster} et~al.}{2014}]{Moster14}
{Moster} B.~P.,  {Macci{\`o}} A.~V.,    {Somerville} R.~S.,  2014, \mnras, 437,
  1027

\bibitem[\protect\citeauthoryear{{Moster}, {Macci{\`o}}, {Somerville}, {Naab}
  \& {Cox}}{{Moster} et~al.}{2011}]{Moster11}
{Moster} B.~P.,  {Macci{\`o}} A.~V.,  {Somerville} R.~S.,  {Naab} T.,    {Cox}
  T.~J.,  2011, \mnras, 415, 3750

\bibitem[\protect\citeauthoryear{{Porter}, {Somerville}, {Primack} \&
  {Johansson}}{{Porter} et~al.}{2014}]{Porter14}
{Porter} L.~A.,  {Somerville} R.~S.,  {Primack} J.~R.,    {Johansson} P.~H.,
  2014, \mnras, 444, 942

\bibitem[\protect\citeauthoryear{{Robertson}, {Bullock}, {Cox}, {Di Matteo},
  {Hernquist}, {Springel} \& {Yoshida}}{{Robertson} et~al.}{2006}]{Robertson06}
{Robertson} B.,  {Bullock} J.~S.,  {Cox} T.~J.,  {Di Matteo} T.,  {Hernquist}
  L.,  {Springel} V.,    {Yoshida} N.,  2006, \apj, 645, 986

\bibitem[\protect\citeauthoryear{{Simard}, {Mendel}, {Patton}, {Ellison} \&
  {McConnachie}}{{Simard} et~al.}{2011}]{Simard11}
{Simard} L.,  {Mendel} J.~T.,  {Patton} D.~R.,  {Ellison} S.~L.,
  {McConnachie} A.~W.,  2011, \apjs, 196, 11

\bibitem[\protect\citeauthoryear{{Somerville}, {Hopkins}, {Cox}, {Robertson} \&
  {Hernquist}}{{Somerville} et~al.}{2008}]{Somerville08}
{Somerville} R.~S.,  {Hopkins} P.~F.,  {Cox} T.~J.,  {Robertson} B.~E.,
  {Hernquist} L.,  2008, \mnras, 391, 481

\bibitem[\protect\citeauthoryear{{Springel}, {White}, {Jenkins}, {Frenk},
  {Yoshida}, {Gao}, {Navarro}, {Thacker}, {Croton}, {Helly}, {Peacock}, {Cole},
  {Thomas}, {Couchman}, {Evrard}, {Colberg} \& {Pearce}}{{Springel}
  et~al.}{2005}]{Springel05}
{Springel} V.,  {White} S.~D.~M.,  {Jenkins} A.,  {Frenk} C.~S.,  {Yoshida} N.,
   {Gao} L.,  {Navarro} J.,  {Thacker} R.,  {Croton} D.,  {Helly} J.,
  {Peacock} J.~A.,  {Cole} S.,  {Thomas} P.,  {Couchman} H.,  {Evrard} A.,
  {Colberg} J.,    {Pearce} F.,  2005, \nat, 435, 629

\bibitem[\protect\citeauthoryear{{Taffoni}, {Mayer}, {Colpi} \&
  {Governato}}{{Taffoni} et~al.}{2003}]{Taffoni03}
{Taffoni} G.,  {Mayer} L.,  {Colpi} M.,    {Governato} F.,  2003, \mnras, 341,
  434

\bibitem[\protect\citeauthoryear{{Villalobos}, {De Lucia} \&
  {Murante}}{{Villalobos} et~al.}{2014}]{Villalobos14}
{Villalobos} {\'A}.,  {De Lucia} G.,    {Murante} G.,  2014, \mnras, 444, 313

\bibitem[\protect\citeauthoryear{{Vulcani}, {Poggianti},
  {Arag{\'o}n-Salamanca}, {Fasano}, {Rudnick}, {Valentinuzzi}, {Dressler},
  {Bettoni}, {Cava}, {D'Onofrio}, {Fritz}, {Moretti}, {Omizzolo} \&
  {Varela}}{{Vulcani} et~al.}{2011}]{Vulcani11}
{Vulcani} B.,  {Poggianti} B.~M.,  {Arag{\'o}n-Salamanca} A.,  {Fasano} G.,
  {Rudnick} G.,  {Valentinuzzi} T.,  {Dressler} A.,  {Bettoni} D.,  {Cava} A.,
  {D'Onofrio} M.,  {Fritz} J.,  {Moretti} A.,  {Omizzolo} A.,    {Varela} J.,
  2011, \mnras, 412, 246

\bibitem[\protect\citeauthoryear{{Wang}, {De Lucia}, {Kitzbichler} \&
  {White}}{{Wang} et~al.}{2008}]{Wang08}
{Wang} J.,  {De Lucia} G.,  {Kitzbichler} M.~G.,    {White} S.~D.~M.,  2008,
  \mnras, 384, 1301

\bibitem[\protect\citeauthoryear{{Weinzirl}, {Jogee}, {Khochfar}, {Burkert} \&
  {Kormendy}}{{Weinzirl} et~al.}{2009}]{Weinzirl09}
{Weinzirl} T.,  {Jogee} S.,  {Khochfar} S.,  {Burkert} A.,    {Kormendy} J.,
  2009, \apj, 696, 411

\bibitem[\protect\citeauthoryear{{Wilman} \& {Erwin}}{{Wilman} \&
  {Erwin}}{2012}]{WilmanErwin12}
{Wilman} D.~J.,  {Erwin} P.,  2012, \apj, 746, 160

\bibitem[\protect\citeauthoryear{{Wilman}, {Fontanot}, {De Lucia}, {Erwin} \&
  {Monaco}}{{Wilman} et~al.}{2013}]{Wilman13}
{Wilman} D.~J.,  {Fontanot} F.,  {De Lucia} G.,  {Erwin} P.,    {Monaco} P.,
  2013, \mnras, 433, 2986

\bibitem[\protect\citeauthoryear{{Wilman}, {Oemler} Jr., {Mulchaey}, {McGee},
  {Balogh} \& {Bower}}{{Wilman} et~al.}{2009}]{Wilman09}
{Wilman} D.~J.,  {Oemler} Jr. A.,  {Mulchaey} J.~S.,  {McGee} S.~L.,  {Balogh}
  M.~L.,    {Bower} R.~G.,  2009, \apj, 692, 298

\bibitem[\protect\citeauthoryear{{Zibetti}}{{Zibetti}}{2008}]{Zibetti08}
{Zibetti} S.,  2008, in {Davies} J.~I.,  {Disney} M.~J.,  eds, IAU Symposium
  Vol.~244 of IAU Symposium, {Statistical Properties of the IntraCluster Light
  from SDSS Image Stacking}.
pp 176--185

\end{thebibliography}

\end{document}